\documentclass[%
 aip,
 amsmath,amssymb,
 reprint,%
]{revtex4-1}

 \usepackage{graphicx}
 \usepackage{dcolumn}
 \usepackage{bm}

\makeatletter
\def\@email#1#2{%
 \endgroup
 \patchcmd{\titleblock@produce}
  {\frontmatter@RRAPformat}
  {\frontmatter@RRAPformat{\produce@RRAP{*#1\href{mailto:#2}{#2}}}\frontmatter@RRAPformat}
  {}{}
}%
\makeatother

\usepackage[auto]{contour}
\usepackage{tikz,pgfplots}
\usetikzlibrary{patterns}
\usepackage{booktabs}
\usepackage{caption}
\usepackage{subcaption}

\usepackage{hyperref}
\hypersetup{
    colorlinks=true,
    linkcolor=blue,
    filecolor=blue,      
    urlcolor=blue,
    citecolor=blue,
    pdfpagemode=FullScreen,
    }

\definecolor{AliceBlue}{rgb}{0.94,0.97,1.00}
\definecolor{AntiqueWhite1}{rgb}{1.00,0.94,0.86}
\definecolor{AntiqueWhite2}{rgb}{0.93,0.87,0.80}
\definecolor{AntiqueWhite3}{rgb}{0.80,0.75,0.69}
\definecolor{AntiqueWhite4}{rgb}{0.55,0.51,0.47}
\definecolor{AntiqueWhite}{rgb}{0.98,0.92,0.84}
\definecolor{BlanchedAlmond}{rgb}{1.00,0.92,0.80}
\definecolor{BlueViolet}{rgb}{0.54,0.17,0.89}
\definecolor{CadetBlue1}{rgb}{0.60,0.96,1.00}
\definecolor{CadetBlue2}{rgb}{0.56,0.90,0.93}
\definecolor{CadetBlue3}{rgb}{0.48,0.77,0.80}
\definecolor{CadetBlue4}{rgb}{0.33,0.53,0.55}
\definecolor{CadetBlue}{rgb}{0.37,0.62,0.63}
\definecolor{CornflowerBlue}{rgb}{0.39,0.58,0.93}
\definecolor{DarkBlue}{rgb}{0.00,0.00,0.55}
\definecolor{DarkCyan}{rgb}{0.00,0.55,0.55}
\definecolor{DarkGoldenrod1}{rgb}{1.00,0.73,0.06}
\definecolor{DarkGoldenrod2}{rgb}{0.93,0.68,0.05}
\definecolor{DarkGoldenrod3}{rgb}{0.80,0.58,0.05}
\definecolor{DarkGoldenrod4}{rgb}{0.55,0.40,0.03}
\definecolor{DarkGoldenrod}{rgb}{0.72,0.53,0.04}
\definecolor{DarkGray}{rgb}{0.66,0.66,0.66}
\definecolor{DarkGreen}{rgb}{0.00,0.39,0.00}
\definecolor{DarkGrey}{rgb}{0.66,0.66,0.66}
\definecolor{DarkKhaki}{rgb}{0.74,0.72,0.42}
\definecolor{DarkMagenta}{rgb}{0.55,0.00,0.55}
\definecolor{DarkOliveGreen1}{rgb}{0.79,1.00,0.44}
\definecolor{DarkOliveGreen2}{rgb}{0.74,0.93,0.41}
\definecolor{DarkOliveGreen3}{rgb}{0.64,0.80,0.35}
\definecolor{DarkOliveGreen4}{rgb}{0.43,0.55,0.24}
\definecolor{DarkOliveGreen}{rgb}{0.33,0.42,0.18}
\definecolor{DarkOrange1}{rgb}{1.00,0.50,0.00}
\definecolor{DarkOrange2}{rgb}{0.93,0.46,0.00}
\definecolor{DarkOrange3}{rgb}{0.80,0.40,0.00}
\definecolor{DarkOrange4}{rgb}{0.55,0.27,0.00}
\definecolor{DarkOrange}{rgb}{1.00,0.55,0.00}
\definecolor{DarkOrchid1}{rgb}{0.75,0.24,1.00}
\definecolor{DarkOrchid2}{rgb}{0.70,0.23,0.93}
\definecolor{DarkOrchid3}{rgb}{0.60,0.20,0.80}
\definecolor{DarkOrchid4}{rgb}{0.41,0.13,0.55}
\definecolor{DarkOrchid}{rgb}{0.60,0.20,0.80}
\definecolor{DarkRed}{rgb}{0.55,0.00,0.00}
\definecolor{DarkSalmon}{rgb}{0.91,0.59,0.48}
\definecolor{DarkSeaGreen1}{rgb}{0.76,1.00,0.76}
\definecolor{DarkSeaGreen2}{rgb}{0.71,0.93,0.71}
\definecolor{DarkSeaGreen3}{rgb}{0.61,0.80,0.61}
\definecolor{DarkSeaGreen4}{rgb}{0.41,0.55,0.41}
\definecolor{DarkSeaGreen}{rgb}{0.56,0.74,0.56}
\definecolor{DarkSlateBlue}{rgb}{0.28,0.24,0.55}
\definecolor{DarkSlateGray1}{rgb}{0.59,1.00,1.00}
\definecolor{DarkSlateGray2}{rgb}{0.55,0.93,0.93}
\definecolor{DarkSlateGray3}{rgb}{0.47,0.80,0.80}
\definecolor{DarkSlateGray4}{rgb}{0.32,0.55,0.55}
\definecolor{DarkSlateGray}{rgb}{0.18,0.31,0.31}
\definecolor{DarkSlateGrey}{rgb}{0.18,0.31,0.31}
\definecolor{DarkTurquoise}{rgb}{0.00,0.81,0.82}
\definecolor{DarkViolet}{rgb}{0.58,0.00,0.83}
\definecolor{DeepPink1}{rgb}{1.00,0.08,0.58}
\definecolor{DeepPink2}{rgb}{0.93,0.07,0.54}
\definecolor{DeepPink3}{rgb}{0.80,0.06,0.46}
\definecolor{DeepPink4}{rgb}{0.55,0.04,0.31}
\definecolor{DeepPink}{rgb}{1.00,0.08,0.58}
\definecolor{DeepSkyBlue1}{rgb}{0.00,0.75,1.00}
\definecolor{DeepSkyBlue2}{rgb}{0.00,0.70,0.93}
\definecolor{DeepSkyBlue3}{rgb}{0.00,0.60,0.80}
\definecolor{DeepSkyBlue4}{rgb}{0.00,0.41,0.55}
\definecolor{DeepSkyBlue}{rgb}{0.00,0.75,1.00}
\definecolor{DimGray}{rgb}{0.41,0.41,0.41}
\definecolor{DimGrey}{rgb}{0.41,0.41,0.41}
\definecolor{DodgerBlue1}{rgb}{0.12,0.56,1.00}
\definecolor{DodgerBlue2}{rgb}{0.11,0.53,0.93}
\definecolor{DodgerBlue3}{rgb}{0.09,0.45,0.80}
\definecolor{DodgerBlue4}{rgb}{0.06,0.31,0.55}
\definecolor{DodgerBlue}{rgb}{0.12,0.56,1.00}
\definecolor{FloralWhite}{rgb}{1.00,0.98,0.94}
\definecolor{ForestGreen}{rgb}{0.13,0.55,0.13}
\definecolor{GhostWhite}{rgb}{0.97,0.97,1.00}
\definecolor{GreenYellow}{rgb}{0.68,1.00,0.18}
\definecolor{HotPink1}{rgb}{1.00,0.43,0.71}
\definecolor{HotPink2}{rgb}{0.93,0.42,0.65}
\definecolor{HotPink3}{rgb}{0.80,0.38,0.56}
\definecolor{HotPink4}{rgb}{0.55,0.23,0.38}
\definecolor{HotPink}{rgb}{1.00,0.41,0.71}
\definecolor{IndianRed1}{rgb}{1.00,0.42,0.42}
\definecolor{IndianRed2}{rgb}{0.93,0.39,0.39}
\definecolor{IndianRed3}{rgb}{0.80,0.33,0.33}
\definecolor{IndianRed4}{rgb}{0.55,0.23,0.23}
\definecolor{IndianRed}{rgb}{0.80,0.36,0.36}
\definecolor{LavenderBlush1}{rgb}{1.00,0.94,0.96}
\definecolor{LavenderBlush2}{rgb}{0.93,0.88,0.90}
\definecolor{LavenderBlush3}{rgb}{0.80,0.76,0.77}
\definecolor{LavenderBlush4}{rgb}{0.55,0.51,0.53}
\definecolor{LavenderBlush}{rgb}{1.00,0.94,0.96}
\definecolor{LawnGreen}{rgb}{0.49,0.99,0.00}
\definecolor{LemonChiffon1}{rgb}{1.00,0.98,0.80}
\definecolor{LemonChiffon2}{rgb}{0.93,0.91,0.75}
\definecolor{LemonChiffon3}{rgb}{0.80,0.79,0.65}
\definecolor{LemonChiffon4}{rgb}{0.55,0.54,0.44}
\definecolor{LemonChiffon}{rgb}{1.00,0.98,0.80}
\definecolor{LightBlue1}{rgb}{0.75,0.94,1.00}
\definecolor{LightBlue2}{rgb}{0.70,0.87,0.93}
\definecolor{LightBlue3}{rgb}{0.60,0.75,0.80}
\definecolor{LightBlue4}{rgb}{0.41,0.51,0.55}
\definecolor{LightBlue}{rgb}{0.68,0.85,0.90}
\definecolor{LightCoral}{rgb}{0.94,0.50,0.50}
\definecolor{LightCyan1}{rgb}{0.88,1.00,1.00}
\definecolor{LightCyan2}{rgb}{0.82,0.93,0.93}
\definecolor{LightCyan3}{rgb}{0.71,0.80,0.80}
\definecolor{LightCyan4}{rgb}{0.48,0.55,0.55}
\definecolor{LightCyan}{rgb}{0.88,1.00,1.00}
\definecolor{LightGoldenrod1}{rgb}{1.00,0.93,0.55}
\definecolor{LightGoldenrod2}{rgb}{0.93,0.86,0.51}
\definecolor{LightGoldenrod3}{rgb}{0.80,0.75,0.44}
\definecolor{LightGoldenrod4}{rgb}{0.55,0.51,0.30}
\definecolor{LightGoldenrodYellow}{rgb}{0.98,0.98,0.82}
\definecolor{LightGoldenrod}{rgb}{0.93,0.87,0.51}
\definecolor{LightGray}{rgb}{0.83,0.83,0.83}
\definecolor{LightGreen}{rgb}{0.56,0.93,0.56}
\definecolor{LightGrey}{rgb}{0.83,0.83,0.83}
\definecolor{LightPink1}{rgb}{1.00,0.68,0.73}
\definecolor{LightPink2}{rgb}{0.93,0.64,0.68}
\definecolor{LightPink3}{rgb}{0.80,0.55,0.58}
\definecolor{LightPink4}{rgb}{0.55,0.37,0.40}
\definecolor{LightPink}{rgb}{1.00,0.71,0.76}
\definecolor{LightSalmon1}{rgb}{1.00,0.63,0.48}
\definecolor{LightSalmon2}{rgb}{0.93,0.58,0.45}
\definecolor{LightSalmon3}{rgb}{0.80,0.51,0.38}
\definecolor{LightSalmon4}{rgb}{0.55,0.34,0.26}
\definecolor{LightSalmon}{rgb}{1.00,0.63,0.48}
\definecolor{LightSeaGreen}{rgb}{0.13,0.70,0.67}
\definecolor{LightSkyBlue1}{rgb}{0.69,0.89,1.00}
\definecolor{LightSkyBlue2}{rgb}{0.64,0.83,0.93}
\definecolor{LightSkyBlue3}{rgb}{0.55,0.71,0.80}
\definecolor{LightSkyBlue4}{rgb}{0.38,0.48,0.55}
\definecolor{LightSkyBlue}{rgb}{0.53,0.81,0.98}
\definecolor{LightSlateBlue}{rgb}{0.52,0.44,1.00}
\definecolor{LightSlateGray}{rgb}{0.47,0.53,0.60}
\definecolor{LightSlateGrey}{rgb}{0.47,0.53,0.60}
\definecolor{LightSteelBlue1}{rgb}{0.79,0.88,1.00}
\definecolor{LightSteelBlue2}{rgb}{0.74,0.82,0.93}
\definecolor{LightSteelBlue3}{rgb}{0.64,0.71,0.80}
\definecolor{LightSteelBlue4}{rgb}{0.43,0.48,0.55}
\definecolor{LightSteelBlue}{rgb}{0.69,0.77,0.87}
\definecolor{LightYellow1}{rgb}{1.00,1.00,0.88}
\definecolor{LightYellow2}{rgb}{0.93,0.93,0.82}
\definecolor{LightYellow3}{rgb}{0.80,0.80,0.71}
\definecolor{LightYellow4}{rgb}{0.55,0.55,0.48}
\definecolor{LightYellow}{rgb}{1.00,1.00,0.88}
\definecolor{LimeGreen}{rgb}{0.20,0.80,0.20}
\definecolor{MediumAquamarine}{rgb}{0.40,0.80,0.67}
\definecolor{MediumBlue}{rgb}{0.00,0.00,0.80}
\definecolor{MediumOrchid1}{rgb}{0.88,0.40,1.00}
\definecolor{MediumOrchid2}{rgb}{0.82,0.37,0.93}
\definecolor{MediumOrchid3}{rgb}{0.71,0.32,0.80}
\definecolor{MediumOrchid4}{rgb}{0.48,0.22,0.55}
\definecolor{MediumOrchid}{rgb}{0.73,0.33,0.83}
\definecolor{MediumPurple1}{rgb}{0.67,0.51,1.00}
\definecolor{MediumPurple2}{rgb}{0.62,0.47,0.93}
\definecolor{MediumPurple3}{rgb}{0.54,0.41,0.80}
\definecolor{MediumPurple4}{rgb}{0.36,0.28,0.55}
\definecolor{MediumPurple}{rgb}{0.58,0.44,0.86}
\definecolor{MediumSeaGreen}{rgb}{0.24,0.70,0.44}
\definecolor{MediumSlateBlue}{rgb}{0.48,0.41,0.93}
\definecolor{MediumSpringGreen}{rgb}{0.00,0.98,0.60}
\definecolor{MediumTurquoise}{rgb}{0.28,0.82,0.80}
\definecolor{MediumVioletRed}{rgb}{0.78,0.08,0.52}
\definecolor{MidnightBlue}{rgb}{0.10,0.10,0.44}
\definecolor{MintCream}{rgb}{0.96,1.00,0.98}
\definecolor{MistyRose1}{rgb}{1.00,0.89,0.88}
\definecolor{MistyRose2}{rgb}{0.93,0.84,0.82}
\definecolor{MistyRose3}{rgb}{0.80,0.72,0.71}
\definecolor{MistyRose4}{rgb}{0.55,0.49,0.48}
\definecolor{MistyRose}{rgb}{1.00,0.89,0.88}
\definecolor{NavajoWhite1}{rgb}{1.00,0.87,0.68}
\definecolor{NavajoWhite2}{rgb}{0.93,0.81,0.63}
\definecolor{NavajoWhite3}{rgb}{0.80,0.70,0.55}
\definecolor{NavajoWhite4}{rgb}{0.55,0.47,0.37}
\definecolor{NavajoWhite}{rgb}{1.00,0.87,0.68}
\definecolor{NavyBlue}{rgb}{0.00,0.00,0.50}
\definecolor{OldLace}{rgb}{0.99,0.96,0.90}
\definecolor{OliveDrab1}{rgb}{0.75,1.00,0.24}
\definecolor{OliveDrab2}{rgb}{0.70,0.93,0.23}
\definecolor{OliveDrab3}{rgb}{0.60,0.80,0.20}
\definecolor{OliveDrab4}{rgb}{0.41,0.55,0.13}
\definecolor{OliveDrab}{rgb}{0.42,0.56,0.14}
\definecolor{OrangeRed1}{rgb}{1.00,0.27,0.00}
\definecolor{OrangeRed2}{rgb}{0.93,0.25,0.00}
\definecolor{OrangeRed3}{rgb}{0.80,0.22,0.00}
\definecolor{OrangeRed4}{rgb}{0.55,0.15,0.00}
\definecolor{OrangeRed}{rgb}{1.00,0.27,0.00}
\definecolor{PaleGoldenrod}{rgb}{0.93,0.91,0.67}
\definecolor{PaleGreen1}{rgb}{0.60,1.00,0.60}
\definecolor{PaleGreen2}{rgb}{0.56,0.93,0.56}
\definecolor{PaleGreen3}{rgb}{0.49,0.80,0.49}
\definecolor{PaleGreen4}{rgb}{0.33,0.55,0.33}
\definecolor{PaleGreen}{rgb}{0.60,0.98,0.60}
\definecolor{PaleTurquoise1}{rgb}{0.73,1.00,1.00}
\definecolor{PaleTurquoise2}{rgb}{0.68,0.93,0.93}
\definecolor{PaleTurquoise3}{rgb}{0.59,0.80,0.80}
\definecolor{PaleTurquoise4}{rgb}{0.40,0.55,0.55}
\definecolor{PaleTurquoise}{rgb}{0.69,0.93,0.93}
\definecolor{PaleVioletRed1}{rgb}{1.00,0.51,0.67}
\definecolor{PaleVioletRed2}{rgb}{0.93,0.47,0.62}
\definecolor{PaleVioletRed3}{rgb}{0.80,0.41,0.54}
\definecolor{PaleVioletRed4}{rgb}{0.55,0.28,0.36}
\definecolor{PaleVioletRed}{rgb}{0.86,0.44,0.58}
\definecolor{PapayaWhip}{rgb}{1.00,0.94,0.84}
\definecolor{PeachPuff1}{rgb}{1.00,0.85,0.73}
\definecolor{PeachPuff2}{rgb}{0.93,0.80,0.68}
\definecolor{PeachPuff3}{rgb}{0.80,0.69,0.58}
\definecolor{PeachPuff4}{rgb}{0.55,0.47,0.40}
\definecolor{PeachPuff}{rgb}{1.00,0.85,0.73}
\definecolor{PowderBlue}{rgb}{0.69,0.88,0.90}
\definecolor{RosyBrown1}{rgb}{1.00,0.76,0.76}
\definecolor{RosyBrown2}{rgb}{0.93,0.71,0.71}
\definecolor{RosyBrown3}{rgb}{0.80,0.61,0.61}
\definecolor{RosyBrown4}{rgb}{0.55,0.41,0.41}
\definecolor{RosyBrown}{rgb}{0.74,0.56,0.56}
\definecolor{RoyalBlue1}{rgb}{0.28,0.46,1.00}
\definecolor{RoyalBlue2}{rgb}{0.26,0.43,0.93}
\definecolor{RoyalBlue3}{rgb}{0.23,0.37,0.80}
\definecolor{RoyalBlue4}{rgb}{0.15,0.25,0.55}
\definecolor{RoyalBlue}{rgb}{0.25,0.41,0.88}
\definecolor{SaddleBrown}{rgb}{0.55,0.27,0.07}
\definecolor{SandyBrown}{rgb}{0.96,0.64,0.38}
\definecolor{SeaGreen1}{rgb}{0.33,1.00,0.62}
\definecolor{SeaGreen2}{rgb}{0.31,0.93,0.58}
\definecolor{SeaGreen3}{rgb}{0.26,0.80,0.50}
\definecolor{SeaGreen4}{rgb}{0.18,0.55,0.34}
\definecolor{SeaGreen}{rgb}{0.18,0.55,0.34}
\definecolor{SkyBlue1}{rgb}{0.53,0.81,1.00}
\definecolor{SkyBlue2}{rgb}{0.49,0.75,0.93}
\definecolor{SkyBlue3}{rgb}{0.42,0.65,0.80}
\definecolor{SkyBlue4}{rgb}{0.29,0.44,0.55}
\definecolor{SkyBlue}{rgb}{0.53,0.81,0.92}
\definecolor{SlateBlue1}{rgb}{0.51,0.44,1.00}
\definecolor{SlateBlue2}{rgb}{0.48,0.40,0.93}
\definecolor{SlateBlue3}{rgb}{0.41,0.35,0.80}
\definecolor{SlateBlue4}{rgb}{0.28,0.24,0.55}
\definecolor{SlateBlue}{rgb}{0.42,0.35,0.80}
\definecolor{SlateGray1}{rgb}{0.78,0.89,1.00}
\definecolor{SlateGray2}{rgb}{0.73,0.83,0.93}
\definecolor{SlateGray3}{rgb}{0.62,0.71,0.80}
\definecolor{SlateGray4}{rgb}{0.42,0.48,0.55}
\definecolor{SlateGray}{rgb}{0.44,0.50,0.56}
\definecolor{SlateGrey}{rgb}{0.44,0.50,0.56}
\definecolor{SpringGreen1}{rgb}{0.00,1.00,0.50}
\definecolor{SpringGreen2}{rgb}{0.00,0.93,0.46}
\definecolor{SpringGreen3}{rgb}{0.00,0.80,0.40}
\definecolor{SpringGreen4}{rgb}{0.00,0.55,0.27}
\definecolor{SpringGreen}{rgb}{0.00,1.00,0.50}
\definecolor{SteelBlue1}{rgb}{0.39,0.72,1.00}
\definecolor{SteelBlue2}{rgb}{0.36,0.67,0.93}
\definecolor{SteelBlue3}{rgb}{0.31,0.58,0.80}
\definecolor{SteelBlue4}{rgb}{0.21,0.39,0.55}
\definecolor{SteelBlue}{rgb}{0.27,0.51,0.71}
\definecolor{VioletRed1}{rgb}{1.00,0.24,0.59}
\definecolor{VioletRed2}{rgb}{0.93,0.23,0.55}
\definecolor{VioletRed3}{rgb}{0.80,0.20,0.47}
\definecolor{VioletRed4}{rgb}{0.55,0.13,0.32}
\definecolor{VioletRed}{rgb}{0.82,0.13,0.56}
\definecolor{WhiteSmoke}{rgb}{0.96,0.96,0.96}
\definecolor{YellowGreen}{rgb}{0.60,0.80,0.20}
\definecolor{aliceblue}{rgb}{0.94,0.97,1.00}
\definecolor{antiquewhite}{rgb}{0.98,0.92,0.84}
\definecolor{aquamarine1}{rgb}{0.50,1.00,0.83}
\definecolor{aquamarine2}{rgb}{0.46,0.93,0.78}
\definecolor{aquamarine3}{rgb}{0.40,0.80,0.67}
\definecolor{aquamarine4}{rgb}{0.27,0.55,0.45}
\definecolor{aquamarine}{rgb}{0.50,1.00,0.83}
\definecolor{azure1}{rgb}{0.94,1.00,1.00}
\definecolor{azure2}{rgb}{0.88,0.93,0.93}
\definecolor{azure3}{rgb}{0.76,0.80,0.80}
\definecolor{azure4}{rgb}{0.51,0.55,0.55}
\definecolor{azure}{rgb}{0.94,1.00,1.00}
\definecolor{beige}{rgb}{0.96,0.96,0.86}
\definecolor{bisque1}{rgb}{1.00,0.89,0.77}
\definecolor{bisque2}{rgb}{0.93,0.84,0.72}
\definecolor{bisque3}{rgb}{0.80,0.72,0.62}
\definecolor{bisque4}{rgb}{0.55,0.49,0.42}
\definecolor{bisque}{rgb}{1.00,0.89,0.77}
\definecolor{black}{rgb}{0.00,0.00,0.00}
\definecolor{blanchedalmond}{rgb}{1.00,0.92,0.80}
\definecolor{blue1}{rgb}{0.00,0.00,1.00}
\definecolor{blue2}{rgb}{0.00,0.00,0.93}
\definecolor{blue3}{rgb}{0.00,0.00,0.80}
\definecolor{blue4}{rgb}{0.00,0.00,0.55}
\definecolor{blueviolet}{rgb}{0.54,0.17,0.89}
\definecolor{blue}{rgb}{0.00,0.00,1.00}
\definecolor{brown1}{rgb}{1.00,0.25,0.25}
\definecolor{brown2}{rgb}{0.93,0.23,0.23}
\definecolor{brown3}{rgb}{0.80,0.20,0.20}
\definecolor{brown4}{rgb}{0.55,0.14,0.14}
\definecolor{brown}{rgb}{0.65,0.16,0.16}
\definecolor{burlywood1}{rgb}{1.00,0.83,0.61}
\definecolor{burlywood2}{rgb}{0.93,0.77,0.57}
\definecolor{burlywood3}{rgb}{0.80,0.67,0.49}
\definecolor{burlywood4}{rgb}{0.55,0.45,0.33}
\definecolor{burlywood}{rgb}{0.87,0.72,0.53}
\definecolor{cadetblue}{rgb}{0.37,0.62,0.63}
\definecolor{chartreuse1}{rgb}{0.50,1.00,0.00}
\definecolor{chartreuse2}{rgb}{0.46,0.93,0.00}
\definecolor{chartreuse3}{rgb}{0.40,0.80,0.00}
\definecolor{chartreuse4}{rgb}{0.27,0.55,0.00}
\definecolor{chartreuse}{rgb}{0.50,1.00,0.00}
\definecolor{chocolate1}{rgb}{1.00,0.50,0.14}
\definecolor{chocolate2}{rgb}{0.93,0.46,0.13}
\definecolor{chocolate3}{rgb}{0.80,0.40,0.11}
\definecolor{chocolate4}{rgb}{0.55,0.27,0.07}
\definecolor{chocolate}{rgb}{0.82,0.41,0.12}
\definecolor{coral1}{rgb}{1.00,0.45,0.34}
\definecolor{coral2}{rgb}{0.93,0.42,0.31}
\definecolor{coral3}{rgb}{0.80,0.36,0.27}
\definecolor{coral4}{rgb}{0.55,0.24,0.18}
\definecolor{coral}{rgb}{1.00,0.50,0.31}
\definecolor{cornflowerblue}{rgb}{0.39,0.58,0.93}
\definecolor{cornsilk1}{rgb}{1.00,0.97,0.86}
\definecolor{cornsilk2}{rgb}{0.93,0.91,0.80}
\definecolor{cornsilk3}{rgb}{0.80,0.78,0.69}
\definecolor{cornsilk4}{rgb}{0.55,0.53,0.47}
\definecolor{cornsilk}{rgb}{1.00,0.97,0.86}
\definecolor{cyan1}{rgb}{0.00,1.00,1.00}
\definecolor{cyan2}{rgb}{0.00,0.93,0.93}
\definecolor{cyan3}{rgb}{0.00,0.80,0.80}
\definecolor{cyan4}{rgb}{0.00,0.55,0.55}
\definecolor{cyan}{rgb}{0.00,1.00,1.00}
\definecolor{darkblue}{rgb}{0.00,0.00,0.55}
\definecolor{darkcyan}{rgb}{0.00,0.55,0.55}
\definecolor{darkgoldenrod}{rgb}{0.72,0.53,0.04}
\definecolor{darkgray}{rgb}{0.66,0.66,0.66}
\definecolor{darkgreen}{rgb}{0.00,0.39,0.00}
\definecolor{darkgrey}{rgb}{0.66,0.66,0.66}
\definecolor{darkkhaki}{rgb}{0.74,0.72,0.42}
\definecolor{darkmagenta}{rgb}{0.55,0.00,0.55}
\definecolor{darkolive}{rgb}{0.33,0.42,0.18}
\definecolor{darkorange}{rgb}{1.00,0.55,0.00}
\definecolor{darkorchid}{rgb}{0.60,0.20,0.80}
\definecolor{darkred}{rgb}{0.55,0.00,0.00}
\definecolor{darksalmon}{rgb}{0.91,0.59,0.48}
\definecolor{darksea}{rgb}{0.56,0.74,0.56}
\definecolor{darkslate}{rgb}{0.18,0.31,0.31}
\definecolor{darkslate}{rgb}{0.18,0.31,0.31}
\definecolor{darkslate}{rgb}{0.28,0.24,0.55}
\definecolor{darkturquoise}{rgb}{0.00,0.81,0.82}
\definecolor{darkviolet}{rgb}{0.58,0.00,0.83}
\definecolor{deeppink}{rgb}{1.00,0.08,0.58}
\definecolor{deepsky}{rgb}{0.00,0.75,1.00}
\definecolor{dimgray}{rgb}{0.41,0.41,0.41}
\definecolor{dimgrey}{rgb}{0.41,0.41,0.41}
\definecolor{dodgerblue}{rgb}{0.12,0.56,1.00}
\definecolor{firebrick1}{rgb}{1.00,0.19,0.19}
\definecolor{firebrick2}{rgb}{0.93,0.17,0.17}
\definecolor{firebrick3}{rgb}{0.80,0.15,0.15}
\definecolor{firebrick4}{rgb}{0.55,0.10,0.10}
\definecolor{firebrick}{rgb}{0.70,0.13,0.13}
\definecolor{floralwhite}{rgb}{1.00,0.98,0.94}
\definecolor{forestgreen}{rgb}{0.13,0.55,0.13}
\definecolor{gainsboro}{rgb}{0.86,0.86,0.86}
\definecolor{ghostwhite}{rgb}{0.97,0.97,1.00}
\definecolor{gold1}{rgb}{1.00,0.84,0.00}
\definecolor{gold2}{rgb}{0.93,0.79,0.00}
\definecolor{gold3}{rgb}{0.80,0.68,0.00}
\definecolor{gold4}{rgb}{0.55,0.46,0.00}
\definecolor{goldenrod1}{rgb}{1.00,0.76,0.15}
\definecolor{goldenrod2}{rgb}{0.93,0.71,0.13}
\definecolor{goldenrod3}{rgb}{0.80,0.61,0.11}
\definecolor{goldenrod4}{rgb}{0.55,0.41,0.08}
\definecolor{goldenrod}{rgb}{0.85,0.65,0.13}
\definecolor{gold}{rgb}{1.00,0.84,0.00}
\definecolor{gray0}{rgb}{0.00,0.00,0.00}
\definecolor{gray100}{rgb}{1.00,1.00,1.00}
\definecolor{gray10}{rgb}{0.10,0.10,0.10}
\definecolor{gray11}{rgb}{0.11,0.11,0.11}
\definecolor{gray12}{rgb}{0.12,0.12,0.12}
\definecolor{gray13}{rgb}{0.13,0.13,0.13}
\definecolor{gray14}{rgb}{0.14,0.14,0.14}
\definecolor{gray15}{rgb}{0.15,0.15,0.15}
\definecolor{gray16}{rgb}{0.16,0.16,0.16}
\definecolor{gray17}{rgb}{0.17,0.17,0.17}
\definecolor{gray18}{rgb}{0.18,0.18,0.18}
\definecolor{gray19}{rgb}{0.19,0.19,0.19}
\definecolor{gray1}{rgb}{0.01,0.01,0.01}
\definecolor{gray20}{rgb}{0.20,0.20,0.20}
\definecolor{gray21}{rgb}{0.21,0.21,0.21}
\definecolor{gray22}{rgb}{0.22,0.22,0.22}
\definecolor{gray23}{rgb}{0.23,0.23,0.23}
\definecolor{gray24}{rgb}{0.24,0.24,0.24}
\definecolor{gray25}{rgb}{0.25,0.25,0.25}
\definecolor{gray26}{rgb}{0.26,0.26,0.26}
\definecolor{gray27}{rgb}{0.27,0.27,0.27}
\definecolor{gray28}{rgb}{0.28,0.28,0.28}
\definecolor{gray29}{rgb}{0.29,0.29,0.29}
\definecolor{gray2}{rgb}{0.02,0.02,0.02}
\definecolor{gray30}{rgb}{0.30,0.30,0.30}
\definecolor{gray31}{rgb}{0.31,0.31,0.31}
\definecolor{gray32}{rgb}{0.32,0.32,0.32}
\definecolor{gray33}{rgb}{0.33,0.33,0.33}
\definecolor{gray34}{rgb}{0.34,0.34,0.34}
\definecolor{gray35}{rgb}{0.35,0.35,0.35}
\definecolor{gray36}{rgb}{0.36,0.36,0.36}
\definecolor{gray37}{rgb}{0.37,0.37,0.37}
\definecolor{gray38}{rgb}{0.38,0.38,0.38}
\definecolor{gray39}{rgb}{0.39,0.39,0.39}
\definecolor{gray3}{rgb}{0.03,0.03,0.03}
\definecolor{gray40}{rgb}{0.40,0.40,0.40}
\definecolor{gray41}{rgb}{0.41,0.41,0.41}
\definecolor{gray42}{rgb}{0.42,0.42,0.42}
\definecolor{gray43}{rgb}{0.43,0.43,0.43}
\definecolor{gray44}{rgb}{0.44,0.44,0.44}
\definecolor{gray45}{rgb}{0.45,0.45,0.45}
\definecolor{gray46}{rgb}{0.46,0.46,0.46}
\definecolor{gray47}{rgb}{0.47,0.47,0.47}
\definecolor{gray48}{rgb}{0.48,0.48,0.48}
\definecolor{gray49}{rgb}{0.49,0.49,0.49}
\definecolor{gray4}{rgb}{0.04,0.04,0.04}
\definecolor{gray50}{rgb}{0.50,0.50,0.50}
\definecolor{gray51}{rgb}{0.51,0.51,0.51}
\definecolor{gray52}{rgb}{0.52,0.52,0.52}
\definecolor{gray53}{rgb}{0.53,0.53,0.53}
\definecolor{gray54}{rgb}{0.54,0.54,0.54}
\definecolor{gray55}{rgb}{0.55,0.55,0.55}
\definecolor{gray56}{rgb}{0.56,0.56,0.56}
\definecolor{gray57}{rgb}{0.57,0.57,0.57}
\definecolor{gray58}{rgb}{0.58,0.58,0.58}
\definecolor{gray59}{rgb}{0.59,0.59,0.59}
\definecolor{gray5}{rgb}{0.05,0.05,0.05}
\definecolor{gray60}{rgb}{0.60,0.60,0.60}
\definecolor{gray61}{rgb}{0.61,0.61,0.61}
\definecolor{gray62}{rgb}{0.62,0.62,0.62}
\definecolor{gray63}{rgb}{0.63,0.63,0.63}
\definecolor{gray64}{rgb}{0.64,0.64,0.64}
\definecolor{gray65}{rgb}{0.65,0.65,0.65}
\definecolor{gray66}{rgb}{0.66,0.66,0.66}
\definecolor{gray67}{rgb}{0.67,0.67,0.67}
\definecolor{gray68}{rgb}{0.68,0.68,0.68}
\definecolor{gray69}{rgb}{0.69,0.69,0.69}
\definecolor{gray6}{rgb}{0.06,0.06,0.06}
\definecolor{gray70}{rgb}{0.70,0.70,0.70}
\definecolor{gray71}{rgb}{0.71,0.71,0.71}
\definecolor{gray72}{rgb}{0.72,0.72,0.72}
\definecolor{gray73}{rgb}{0.73,0.73,0.73}
\definecolor{gray74}{rgb}{0.74,0.74,0.74}
\definecolor{gray75}{rgb}{0.75,0.75,0.75}
\definecolor{gray76}{rgb}{0.76,0.76,0.76}
\definecolor{gray77}{rgb}{0.77,0.77,0.77}
\definecolor{gray78}{rgb}{0.78,0.78,0.78}
\definecolor{gray79}{rgb}{0.79,0.79,0.79}
\definecolor{gray7}{rgb}{0.07,0.07,0.07}
\definecolor{gray80}{rgb}{0.80,0.80,0.80}
\definecolor{gray81}{rgb}{0.81,0.81,0.81}
\definecolor{gray82}{rgb}{0.82,0.82,0.82}
\definecolor{gray83}{rgb}{0.83,0.83,0.83}
\definecolor{gray84}{rgb}{0.84,0.84,0.84}
\definecolor{gray85}{rgb}{0.85,0.85,0.85}
\definecolor{gray86}{rgb}{0.86,0.86,0.86}
\definecolor{gray87}{rgb}{0.87,0.87,0.87}
\definecolor{gray88}{rgb}{0.88,0.88,0.88}
\definecolor{gray89}{rgb}{0.89,0.89,0.89}
\definecolor{gray8}{rgb}{0.08,0.08,0.08}
\definecolor{gray90}{rgb}{0.90,0.90,0.90}
\definecolor{gray91}{rgb}{0.91,0.91,0.91}
\definecolor{gray92}{rgb}{0.92,0.92,0.92}
\definecolor{gray93}{rgb}{0.93,0.93,0.93}
\definecolor{gray94}{rgb}{0.94,0.94,0.94}
\definecolor{gray95}{rgb}{0.95,0.95,0.95}
\definecolor{gray96}{rgb}{0.96,0.96,0.96}
\definecolor{gray97}{rgb}{0.97,0.97,0.97}
\definecolor{gray98}{rgb}{0.98,0.98,0.98}
\definecolor{gray99}{rgb}{0.99,0.99,0.99}
\definecolor{gray9}{rgb}{0.09,0.09,0.09}
\definecolor{gray}{rgb}{0.75,0.75,0.75}
\definecolor{green1}{rgb}{0.00,1.00,0.00}
\definecolor{green2}{rgb}{0.00,0.93,0.00}
\definecolor{green3}{rgb}{0.00,0.80,0.00}
\definecolor{green4}{rgb}{0.00,0.55,0.00}
\definecolor{greenyellow}{rgb}{0.68,1.00,0.18}
\definecolor{green}{rgb}{0.00,1.00,0.00}
\definecolor{grey0}{rgb}{0.00,0.00,0.00}
\definecolor{grey100}{rgb}{1.00,1.00,1.00}
\definecolor{grey10}{rgb}{0.10,0.10,0.10}
\definecolor{grey11}{rgb}{0.11,0.11,0.11}
\definecolor{grey12}{rgb}{0.12,0.12,0.12}
\definecolor{grey13}{rgb}{0.13,0.13,0.13}
\definecolor{grey14}{rgb}{0.14,0.14,0.14}
\definecolor{grey15}{rgb}{0.15,0.15,0.15}
\definecolor{grey16}{rgb}{0.16,0.16,0.16}
\definecolor{grey17}{rgb}{0.17,0.17,0.17}
\definecolor{grey18}{rgb}{0.18,0.18,0.18}
\definecolor{grey19}{rgb}{0.19,0.19,0.19}
\definecolor{grey1}{rgb}{0.01,0.01,0.01}
\definecolor{grey20}{rgb}{0.20,0.20,0.20}
\definecolor{grey21}{rgb}{0.21,0.21,0.21}
\definecolor{grey22}{rgb}{0.22,0.22,0.22}
\definecolor{grey23}{rgb}{0.23,0.23,0.23}
\definecolor{grey24}{rgb}{0.24,0.24,0.24}
\definecolor{grey25}{rgb}{0.25,0.25,0.25}
\definecolor{grey26}{rgb}{0.26,0.26,0.26}
\definecolor{grey27}{rgb}{0.27,0.27,0.27}
\definecolor{grey28}{rgb}{0.28,0.28,0.28}
\definecolor{grey29}{rgb}{0.29,0.29,0.29}
\definecolor{grey2}{rgb}{0.02,0.02,0.02}
\definecolor{grey30}{rgb}{0.30,0.30,0.30}
\definecolor{grey31}{rgb}{0.31,0.31,0.31}
\definecolor{grey32}{rgb}{0.32,0.32,0.32}
\definecolor{grey33}{rgb}{0.33,0.33,0.33}
\definecolor{grey34}{rgb}{0.34,0.34,0.34}
\definecolor{grey35}{rgb}{0.35,0.35,0.35}
\definecolor{grey36}{rgb}{0.36,0.36,0.36}
\definecolor{grey37}{rgb}{0.37,0.37,0.37}
\definecolor{grey38}{rgb}{0.38,0.38,0.38}
\definecolor{grey39}{rgb}{0.39,0.39,0.39}
\definecolor{grey3}{rgb}{0.03,0.03,0.03}
\definecolor{grey40}{rgb}{0.40,0.40,0.40}
\definecolor{grey41}{rgb}{0.41,0.41,0.41}
\definecolor{grey42}{rgb}{0.42,0.42,0.42}
\definecolor{grey43}{rgb}{0.43,0.43,0.43}
\definecolor{grey44}{rgb}{0.44,0.44,0.44}
\definecolor{grey45}{rgb}{0.45,0.45,0.45}
\definecolor{grey46}{rgb}{0.46,0.46,0.46}
\definecolor{grey47}{rgb}{0.47,0.47,0.47}
\definecolor{grey48}{rgb}{0.48,0.48,0.48}
\definecolor{grey49}{rgb}{0.49,0.49,0.49}
\definecolor{grey4}{rgb}{0.04,0.04,0.04}
\definecolor{grey50}{rgb}{0.50,0.50,0.50}
\definecolor{grey51}{rgb}{0.51,0.51,0.51}
\definecolor{grey52}{rgb}{0.52,0.52,0.52}
\definecolor{grey53}{rgb}{0.53,0.53,0.53}
\definecolor{grey54}{rgb}{0.54,0.54,0.54}
\definecolor{grey55}{rgb}{0.55,0.55,0.55}
\definecolor{grey56}{rgb}{0.56,0.56,0.56}
\definecolor{grey57}{rgb}{0.57,0.57,0.57}
\definecolor{grey58}{rgb}{0.58,0.58,0.58}
\definecolor{grey59}{rgb}{0.59,0.59,0.59}
\definecolor{grey5}{rgb}{0.05,0.05,0.05}
\definecolor{grey60}{rgb}{0.60,0.60,0.60}
\definecolor{grey61}{rgb}{0.61,0.61,0.61}
\definecolor{grey62}{rgb}{0.62,0.62,0.62}
\definecolor{grey63}{rgb}{0.63,0.63,0.63}
\definecolor{grey64}{rgb}{0.64,0.64,0.64}
\definecolor{grey65}{rgb}{0.65,0.65,0.65}
\definecolor{grey66}{rgb}{0.66,0.66,0.66}
\definecolor{grey67}{rgb}{0.67,0.67,0.67}
\definecolor{grey68}{rgb}{0.68,0.68,0.68}
\definecolor{grey69}{rgb}{0.69,0.69,0.69}
\definecolor{grey6}{rgb}{0.06,0.06,0.06}
\definecolor{grey70}{rgb}{0.70,0.70,0.70}
\definecolor{grey71}{rgb}{0.71,0.71,0.71}
\definecolor{grey72}{rgb}{0.72,0.72,0.72}
\definecolor{grey73}{rgb}{0.73,0.73,0.73}
\definecolor{grey74}{rgb}{0.74,0.74,0.74}
\definecolor{grey75}{rgb}{0.75,0.75,0.75}
\definecolor{grey76}{rgb}{0.76,0.76,0.76}
\definecolor{grey77}{rgb}{0.77,0.77,0.77}
\definecolor{grey78}{rgb}{0.78,0.78,0.78}
\definecolor{grey79}{rgb}{0.79,0.79,0.79}
\definecolor{grey7}{rgb}{0.07,0.07,0.07}
\definecolor{grey80}{rgb}{0.80,0.80,0.80}
\definecolor{grey81}{rgb}{0.81,0.81,0.81}
\definecolor{grey82}{rgb}{0.82,0.82,0.82}
\definecolor{grey83}{rgb}{0.83,0.83,0.83}
\definecolor{grey84}{rgb}{0.84,0.84,0.84}
\definecolor{grey85}{rgb}{0.85,0.85,0.85}
\definecolor{grey86}{rgb}{0.86,0.86,0.86}
\definecolor{grey87}{rgb}{0.87,0.87,0.87}
\definecolor{grey88}{rgb}{0.88,0.88,0.88}
\definecolor{grey89}{rgb}{0.89,0.89,0.89}
\definecolor{grey8}{rgb}{0.08,0.08,0.08}
\definecolor{grey90}{rgb}{0.90,0.90,0.90}
\definecolor{grey91}{rgb}{0.91,0.91,0.91}
\definecolor{grey92}{rgb}{0.92,0.92,0.92}
\definecolor{grey93}{rgb}{0.93,0.93,0.93}
\definecolor{grey94}{rgb}{0.94,0.94,0.94}
\definecolor{grey95}{rgb}{0.95,0.95,0.95}
\definecolor{grey96}{rgb}{0.96,0.96,0.96}
\definecolor{grey97}{rgb}{0.97,0.97,0.97}
\definecolor{grey98}{rgb}{0.98,0.98,0.98}
\definecolor{grey99}{rgb}{0.99,0.99,0.99}
\definecolor{grey9}{rgb}{0.09,0.09,0.09}
\definecolor{grey}{rgb}{0.75,0.75,0.75}
\definecolor{honeydew1}{rgb}{0.94,1.00,0.94}
\definecolor{honeydew2}{rgb}{0.88,0.93,0.88}
\definecolor{honeydew3}{rgb}{0.76,0.80,0.76}
\definecolor{honeydew4}{rgb}{0.51,0.55,0.51}
\definecolor{honeydew}{rgb}{0.94,1.00,0.94}
\definecolor{hotpink}{rgb}{1.00,0.41,0.71}
\definecolor{indianred}{rgb}{0.80,0.36,0.36}
\definecolor{ivory1}{rgb}{1.00,1.00,0.94}
\definecolor{ivory2}{rgb}{0.93,0.93,0.88}
\definecolor{ivory3}{rgb}{0.80,0.80,0.76}
\definecolor{ivory4}{rgb}{0.55,0.55,0.51}
\definecolor{ivory}{rgb}{1.00,1.00,0.94}
\definecolor{khaki1}{rgb}{1.00,0.96,0.56}
\definecolor{khaki2}{rgb}{0.93,0.90,0.52}
\definecolor{khaki3}{rgb}{0.80,0.78,0.45}
\definecolor{khaki4}{rgb}{0.55,0.53,0.31}
\definecolor{khaki}{rgb}{0.94,0.90,0.55}
\definecolor{lavenderblush}{rgb}{1.00,0.94,0.96}
\definecolor{lavender}{rgb}{0.90,0.90,0.98}
\definecolor{lawngreen}{rgb}{0.49,0.99,0.00}
\definecolor{lemonchiffon}{rgb}{1.00,0.98,0.80}
\definecolor{lightblue}{rgb}{0.68,0.85,0.90}
\definecolor{lightcoral}{rgb}{0.94,0.50,0.50}
\definecolor{lightcyan}{rgb}{0.88,1.00,1.00}
\definecolor{lightgoldenrod}{rgb}{0.93,0.87,0.51}
\definecolor{lightgoldenrod}{rgb}{0.98,0.98,0.82}
\definecolor{lightgray}{rgb}{0.83,0.83,0.83}
\definecolor{lightgreen}{rgb}{0.56,0.93,0.56}
\definecolor{lightgrey}{rgb}{0.83,0.83,0.83}
\definecolor{lightpink}{rgb}{1.00,0.71,0.76}
\definecolor{lightsalmon}{rgb}{1.00,0.63,0.48}
\definecolor{lightsea}{rgb}{0.13,0.70,0.67}
\definecolor{lightsky}{rgb}{0.53,0.81,0.98}
\definecolor{lightslate}{rgb}{0.47,0.53,0.60}
\definecolor{lightslate}{rgb}{0.47,0.53,0.60}
\definecolor{lightslate}{rgb}{0.52,0.44,1.00}
\definecolor{lightsteel}{rgb}{0.69,0.77,0.87}
\definecolor{lightyellow}{rgb}{1.00,1.00,0.88}
\definecolor{limegreen}{rgb}{0.20,0.80,0.20}
\definecolor{linen}{rgb}{0.98,0.94,0.90}
\definecolor{magenta1}{rgb}{1.00,0.00,1.00}
\definecolor{magenta2}{rgb}{0.93,0.00,0.93}
\definecolor{magenta3}{rgb}{0.80,0.00,0.80}
\definecolor{magenta4}{rgb}{0.55,0.00,0.55}
\definecolor{magenta}{rgb}{1.00,0.00,1.00}
\definecolor{maroon1}{rgb}{1.00,0.20,0.70}
\definecolor{maroon2}{rgb}{0.93,0.19,0.65}
\definecolor{maroon3}{rgb}{0.80,0.16,0.56}
\definecolor{maroon4}{rgb}{0.55,0.11,0.38}
\definecolor{maroon}{rgb}{0.69,0.19,0.38}
\definecolor{mediumaquamarine}{rgb}{0.40,0.80,0.67}
\definecolor{mediumblue}{rgb}{0.00,0.00,0.80}
\definecolor{mediumorchid}{rgb}{0.73,0.33,0.83}
\definecolor{mediumpurple}{rgb}{0.58,0.44,0.86}
\definecolor{mediumsea}{rgb}{0.24,0.70,0.44}
\definecolor{mediumslate}{rgb}{0.48,0.41,0.93}
\definecolor{mediumspring}{rgb}{0.00,0.98,0.60}
\definecolor{mediumturquoise}{rgb}{0.28,0.82,0.80}
\definecolor{mediumviolet}{rgb}{0.78,0.08,0.52}
\definecolor{midnightblue}{rgb}{0.10,0.10,0.44}
\definecolor{mintcream}{rgb}{0.96,1.00,0.98}
\definecolor{mistyrose}{rgb}{1.00,0.89,0.88}
\definecolor{moccasin}{rgb}{1.00,0.89,0.71}
\definecolor{navajowhite}{rgb}{1.00,0.87,0.68}
\definecolor{navyblue}{rgb}{0.00,0.00,0.50}
\definecolor{navy}{rgb}{0.00,0.00,0.50}
\definecolor{oldlace}{rgb}{0.99,0.96,0.90}
\definecolor{olivedrab}{rgb}{0.42,0.56,0.14}
\definecolor{orange1}{rgb}{1.00,0.65,0.00}
\definecolor{orange2}{rgb}{0.93,0.60,0.00}
\definecolor{orange3}{rgb}{0.80,0.52,0.00}
\definecolor{orange4}{rgb}{0.55,0.35,0.00}
\definecolor{orangered}{rgb}{1.00,0.27,0.00}
\definecolor{orange}{rgb}{1.00,0.65,0.00}
\definecolor{orchid1}{rgb}{1.00,0.51,0.98}
\definecolor{orchid2}{rgb}{0.93,0.48,0.91}
\definecolor{orchid3}{rgb}{0.80,0.41,0.79}
\definecolor{orchid4}{rgb}{0.55,0.28,0.54}
\definecolor{orchid}{rgb}{0.85,0.44,0.84}
\definecolor{palegoldenrod}{rgb}{0.93,0.91,0.67}
\definecolor{palegreen}{rgb}{0.60,0.98,0.60}
\definecolor{paleturquoise}{rgb}{0.69,0.93,0.93}
\definecolor{paleviolet}{rgb}{0.86,0.44,0.58}
\definecolor{papayawhip}{rgb}{1.00,0.94,0.84}
\definecolor{peachpuff}{rgb}{1.00,0.85,0.73}
\definecolor{peru}{rgb}{0.80,0.52,0.25}
\definecolor{pink1}{rgb}{1.00,0.71,0.77}
\definecolor{pink2}{rgb}{0.93,0.66,0.72}
\definecolor{pink3}{rgb}{0.80,0.57,0.62}
\definecolor{pink4}{rgb}{0.55,0.39,0.42}
\definecolor{pink}{rgb}{1.00,0.75,0.80}
\definecolor{plum1}{rgb}{1.00,0.73,1.00}
\definecolor{plum2}{rgb}{0.93,0.68,0.93}
\definecolor{plum3}{rgb}{0.80,0.59,0.80}
\definecolor{plum4}{rgb}{0.55,0.40,0.55}
\definecolor{plum}{rgb}{0.87,0.63,0.87}
\definecolor{powderblue}{rgb}{0.69,0.88,0.90}
\definecolor{purple1}{rgb}{0.61,0.19,1.00}
\definecolor{purple2}{rgb}{0.57,0.17,0.93}
\definecolor{purple3}{rgb}{0.49,0.15,0.80}
\definecolor{purple4}{rgb}{0.33,0.10,0.55}
\definecolor{purple}{rgb}{0.63,0.13,0.94}
\definecolor{red1}{rgb}{1.00,0.00,0.00}
\definecolor{red2}{rgb}{0.93,0.00,0.00}
\definecolor{red3}{rgb}{0.80,0.00,0.00}
\definecolor{red4}{rgb}{0.55,0.00,0.00}
\definecolor{red}{rgb}{1.00,0.00,0.00}
\definecolor{rosybrown}{rgb}{0.74,0.56,0.56}
\definecolor{royalblue}{rgb}{0.25,0.41,0.88}
\definecolor{saddlebrown}{rgb}{0.55,0.27,0.07}
\definecolor{salmon1}{rgb}{1.00,0.55,0.41}
\definecolor{salmon2}{rgb}{0.93,0.51,0.38}
\definecolor{salmon3}{rgb}{0.80,0.44,0.33}
\definecolor{salmon4}{rgb}{0.55,0.30,0.22}
\definecolor{salmon}{rgb}{0.98,0.50,0.45}
\definecolor{sandybrown}{rgb}{0.96,0.64,0.38}
\definecolor{seagreen}{rgb}{0.18,0.55,0.34}
\definecolor{seashell1}{rgb}{1.00,0.96,0.93}
\definecolor{seashell2}{rgb}{0.93,0.90,0.87}
\definecolor{seashell3}{rgb}{0.80,0.77,0.75}
\definecolor{seashell4}{rgb}{0.55,0.53,0.51}
\definecolor{seashell}{rgb}{1.00,0.96,0.93}
\definecolor{sienna1}{rgb}{1.00,0.51,0.28}
\definecolor{sienna2}{rgb}{0.93,0.47,0.26}
\definecolor{sienna3}{rgb}{0.80,0.41,0.22}
\definecolor{sienna4}{rgb}{0.55,0.28,0.15}
\definecolor{sienna}{rgb}{0.63,0.32,0.18}
\definecolor{skyblue}{rgb}{0.53,0.81,0.92}
\definecolor{slateblue}{rgb}{0.42,0.35,0.80}
\definecolor{slategray}{rgb}{0.44,0.50,0.56}
\definecolor{slategrey}{rgb}{0.44,0.50,0.56}
\definecolor{snow1}{rgb}{1.00,0.98,0.98}
\definecolor{snow2}{rgb}{0.93,0.91,0.91}
\definecolor{snow3}{rgb}{0.80,0.79,0.79}
\definecolor{snow4}{rgb}{0.55,0.54,0.54}
\definecolor{snow}{rgb}{1.00,0.98,0.98}
\definecolor{springgreen}{rgb}{0.00,1.00,0.50}
\definecolor{steelblue}{rgb}{0.27,0.51,0.71}
\definecolor{tan1}{rgb}{1.00,0.65,0.31}
\definecolor{tan2}{rgb}{0.93,0.60,0.29}
\definecolor{tan3}{rgb}{0.80,0.52,0.25}
\definecolor{tan4}{rgb}{0.55,0.35,0.17}
\definecolor{tan}{rgb}{0.82,0.71,0.55}
\definecolor{thistle1}{rgb}{1.00,0.88,1.00}
\definecolor{thistle2}{rgb}{0.93,0.82,0.93}
\definecolor{thistle3}{rgb}{0.80,0.71,0.80}
\definecolor{thistle4}{rgb}{0.55,0.48,0.55}
\definecolor{thistle}{rgb}{0.85,0.75,0.85}
\definecolor{tomato1}{rgb}{1.00,0.39,0.28}
\definecolor{tomato2}{rgb}{0.93,0.36,0.26}
\definecolor{tomato3}{rgb}{0.80,0.31,0.22}
\definecolor{tomato4}{rgb}{0.55,0.21,0.15}
\definecolor{tomato}{rgb}{1.00,0.39,0.28}
\definecolor{turquoise1}{rgb}{0.00,0.96,1.00}
\definecolor{turquoise2}{rgb}{0.00,0.90,0.93}
\definecolor{turquoise3}{rgb}{0.00,0.77,0.80}
\definecolor{turquoise4}{rgb}{0.00,0.53,0.55}
\definecolor{turquoise}{rgb}{0.25,0.88,0.82}
\definecolor{violetred}{rgb}{0.82,0.13,0.56}
\definecolor{violet}{rgb}{0.93,0.51,0.93}
\definecolor{wheat1}{rgb}{1.00,0.91,0.73}
\definecolor{wheat2}{rgb}{0.93,0.85,0.68}
\definecolor{wheat3}{rgb}{0.80,0.73,0.59}
\definecolor{wheat4}{rgb}{0.55,0.49,0.40}
\definecolor{wheat}{rgb}{0.96,0.87,0.70}
\definecolor{whitesmoke}{rgb}{0.96,0.96,0.96}
\definecolor{white}{rgb}{1.00,1.00,1.00}
\definecolor{yellow1}{rgb}{1.00,1.00,0.00}
\definecolor{yellow2}{rgb}{0.93,0.93,0.00}
\definecolor{yellow3}{rgb}{0.80,0.80,0.00}
\definecolor{yellow4}{rgb}{0.55,0.55,0.00}
\definecolor{yellowgreen}{rgb}{0.60,0.80,0.20}
\definecolor{yellow}{rgb}{1.00,1.00,0.00}

\definecolor{kit-green100}{rgb}{0,.59,.51}
\definecolor{kit-green70}{rgb}{.3,.71,.65}
\definecolor{kit-green50}{rgb}{.50,.79,.75}
\definecolor{kit-green30}{rgb}{.69,.87,.85}
\definecolor{kit-green15}{rgb}{.85,.93,.93}

\definecolor{kit-blue100}{rgb}{.27,.39,.67}
\definecolor{kit-blue70}{rgb}{.49,.57,.76}
\definecolor{kit-blue50}{rgb}{.64,.69,.83}
\definecolor{kit-blue30}{rgb}{.78,.82,.9}
\definecolor{kit-blue15}{rgb}{.89,.91,.95}

\definecolor{kit-yellow100}{cmyk}{0,.05,1,0}
\definecolor{kit-yellow70}{cmyk}{0,.035,.7,0}
\definecolor{kit-yellow50}{cmyk}{0,.025,.5,0}
\definecolor{kit-yellow30}{cmyk}{0,.015,.3,0}
\definecolor{kit-yellow15}{cmyk}{0,.0075,.15,0}

\definecolor{kit-orange100}{cmyk}{0,.45,1,0}
\definecolor{kit-orange70}{cmyk}{0,.315,.7,0}
\definecolor{kit-orange50}{cmyk}{0,.225,.5,0}
\definecolor{kit-orange30}{cmyk}{0,.135,.3,0}
\definecolor{kit-orange15}{cmyk}{0,.0675,.15,0}

\definecolor{kit-red100}{cmyk}{.25,1,1,0}
\definecolor{kit-red70}{cmyk}{.175,.7,.7,0}
\definecolor{kit-red50}{cmyk}{.125,.5,.5,0}
\definecolor{kit-red30}{cmyk}{.075,.3,.3,0}
\definecolor{kit-red15}{cmyk}{.0375,.15,.15,0}
\pgfplotsset{compat=1.18} 

\begin{document}

\preprint{AIP/123-QED}

\title[]{Machine learning for rapid discovery of laminar flow channel wall modifications that enhance heat transfer}
\author{Yuri Koide}
\affiliation{Institute of Theoretical Informatics, Karlsruhe Institute of Technology, Karlsruhe, Germany}
\author{Arjun J. Kaithakkal}
\affiliation{Institute of Fluid Mechanics, Karlsruhe Institute of Technology, Karlsruhe, Germany}
\author{Matthias Schniewind}
\affiliation{Institute of Theoretical Informatics, Karlsruhe Institute of Technology, Karlsruhe, Germany}
\author{Bradley P. Ladewig}
\affiliation{Institute for Micro Process Engineering, Karlsruhe Institute of Technology, Karlsruhe, Germany}
\author{Alexander Stroh}
\affiliation{Institute of Fluid Mechanics, Karlsruhe Institute of Technology, Karlsruhe, Germany}
\author{Pascal Friederich}
\affiliation{Institute of Theoretical Informatics, Karlsruhe Institute of Technology, Karlsruhe, Germany}
\affiliation{Institute of Nanotechnology, Karlsruhe Institute of Technology, Karlsruhe, Germany}

\date{\today}

\begin{abstract}
 Numerical simulation of fluids plays an essential role in modeling many physical phenomena, which enables technological advancements, contributes to sustainable practices, and expands our understanding of various natural and engineered systems. The calculation of heat transfer in fluid flow in simple flat channels is a relatively easy task for various simulation methods. However, once the channel geometry becomes more complex, numerical simulations become a bottleneck in optimizing wall geometries. We present a combination of accurate numerical simulations of arbitrary, flat, and non-flat channels and machine learning models predicting drag coefficient and Stanton number. We show that convolutional neural networks (CNN) can accurately predict the target properties at a fraction of the time of numerical simulations. We use the CNN models in a virtual high-throughput screening approach to explore a large number of possible, randomly generated wall architectures. 
 Data Augmentation was applied to existing geometries data to add generated new training data which have the same number of parameters of heat transfer to improve the model’s generalization.
 The general approach is not only applicable to simple flow setups as presented here but can be extended to more complex tasks, such as multiphase or even reactive unit operations in chemical engineering.\end{abstract}

\maketitle


\section{Introduction}

Heat transfer in fluid flow is an important physical phenomenon, with relevance across all areas of science and engineering ranging from microfluidic devices in chemical engineering and biomedical implants, all the way to high-temperature physics and cosmology. In this proof-of-concept study, we explore an interesting engineering question, which could be posed as "Is it possible to introduce structural changes to the wall of a channel that increases heat transfer, without a corresponding increase in the pressure drop?". 
This fundamental question linked to the ultimate goal of \textit{dissimilar flow control} or \textit{dissimilar heat transfer enhancement} has been asked for decades by various research groups in different application fields.

Dissimilar heat transfer enhancement is proven to be extremely challenging due to similarity in the mechanisms of momentum and heat transfer~\cite{bejan2013convection}.
Investigations of various surfaces including specially designed fins~\cite{manglik1995heat,kays1984compact}, dimples~\cite{elyyan2008investigation} or vortex generators~\cite{fiebig1995vortex} report that an increase in heat transfer (described by Stanton number $St$) is always accompanied by inevitable manifold increase in the drag coefficient $C_f$, which eventually results in a decrease of the Reynolds analogy factor $RA=2St/C_f$~\cite{reynolds1901extent} in comparison to a flat channel configuration. 
It is, however, known that a dissimilar modification of momentum and heat transfer is possible when more sophisticated flow control methods are applied.
Those control methods, for instance, can be based on the introduction of flow perturbations or optimally distributed blowing/suction profiles from the wall surface~\cite{higashi2011simultaneous,hassanzadeh2014wall,Motoki2018Optimal,kaithakkal2020dissimilarity}.
These studies confirm, that a significant enhancement of the Reynolds analogy factor (tripling $RA$ in comparison to the uncontrolled channel flow) is possible when an appropriate flow manipulation is created.
It is found that an introduction of large-scale spanwise rolls significantly promotes heat transfer while the drag coefficient remains less affected.
This concept has also been successfully tested in the framework of turbulent channel flows, where $RA>2$ can be achieved instead of $RA=1$ in an uncontrolled flow configuration~\cite{hasegawa2011dissimilar,yamamoto2013optimal}.
Recent studies in turbulent flows also report a possibility of $RA$ modification using streamwise elongated structures leading to the formation of turbulence-driven secondary motions~\cite{stroh2020secondary}.
The modification of $RA$ is however limited in this case to several percent due to the increase of the wetted area and the corresponding increase in $C_f$.


To simplify the scenario, we consider a two-dimensional channel with laminar flow, heat transfer, and immersed boundary method for the introduction of surface structuring. 
This allows to quickly execute direct numerical simulations (DNS), where a large set of arbitrarily generated surfaces can be investigated. 
In this proof-of-principle study, we present a workflow consisting of numerical simulations (Section~\ref{sec:numerical_procedure} and Section~\ref{sec:performance_indices}), the generation of a dataset from numerical simulations, and the training of machine learning (ML) models (Section~\ref{sec:dataset_and_ML}).
The “hybrid” approaches can be used to explore heat transfer enhancement efficiently compared to only doing numerical simulations. 
 The idea of using “pure” ML is to replace the entire Navier–Stokes simulation with approximations based on deep neural networks.~\cite{li2020neural,bhattacharya2021model,kim2019deep} 
To utilize this cost-efficient approach, precise calculation of geometries generation for ML training is needed. 
We show that the ML model can predict fluid flow and heat transfer characteristics with a large speedup compared to numerical simulations and with a high enough accuracy to screen a large database of possible channel geometries (Section~\ref{sec:results}). We believe that fluid dynamics is central to transportation, health, and defense systems, and it is, therefore, essential that ML solutions are interpretable, explainable, and generalizable. 


\section{Methods}

\subsection{Numerical procedure}
\label{sec:numerical_procedure}

\begin{figure*}
    \input{tikz/lam_prof_rgh}
    \caption{\label{fig:lamprof3} Laminar channel flow with imposed wall structuring.}
\end{figure*}

For the problem setup, we consider a laminar channel flow with arbitrary wall structuring.
The coordinate system of the numerical domain and its geometry ($L_x \times L_y = 10 \delta \times 2 \delta$ with $\delta$ being the half channel height) are illustrated in Figure ~\ref{fig:lamprof3}, where ($x$, $y$) = ($x_1$, $x_2$) correspond to the streamwise and wall-normal  directions respectively. 
The velocity components in the two directions are denoted by ($u$, $v$) = ($u_1$, $u_2$).
The analysis is carried out using flow and temperature fields produced by DNS in a channel flow driven at a constant flow rate (CFR). 
Assuming an incompressible flow, the velocity field is required to satisfy continuity:
\begin{equation}
{\frac{\partial u_i}{\partial x_i} = 0 ,}
\label{eq:solenoid}
\end{equation}
and the Navier-Stokes equations for a constant property Newtonian fluid:
\begin{equation}
\label{eq:impulsgleichkomplett}
    \frac{\partial u_i}{\partial t} + \frac{\partial u_i u_j}{\partial x_j}
= \frac{1}{\rho} P_x \delta_{i1} -\frac{1}{\rho} \frac {\partial p} {\partial x_i} + \nu \frac {\partial^2 u_i} {\partial {x_j} \partial {x_j}  }  + F_{\textrm{IBM},i}.
\end{equation}
Here $p$ is the fluctuating pressure part, $\rho$ is density, $\nu$ is the kinematic viscosity and $F_{{IBM},i}$ represents the external volume force per unit mass required for the immersed boundary method (IBM) with which the structured surface is introduced into the flow domain \cite{Goldstein_1993}.
In the present configuration $F_{\textrm{IBM},i}$ corresponds to the frictional drag between the flow and the part of the surface reproduced by the immersed boundary method, i.e. the structured wall surface.
$P_x$ is the absolute value of the mean streamwise pressure gradient added to the equation in order to drive the flow through the channel.

Due to the CFR approach the bulk Reynolds number is fixed to $\mathrm{Re}_b= 2 U_b \delta / \nu = 200$
for all considered simulations, where $U_b$ is the bulk mean velocity. This means that any modification of the flow is translated into an alteration of the resulting mean streamwise pressure gradient $P_x$ required to maintain the chosen flow rate.
Periodic boundary conditions are applied in the streamwise directions while the wall-normal extension of the flow domain is bounded by no-slip boundary conditions at the lower and upper domain wall ($y = 0,2 \delta$). Subscript $l$ and $u$ are used throughout the manuscript to denote quantities on the lower and upper walls, respectively.

Temperature $T$ is treated as a passive scalar and has to satisfy the scalar transport equation: 
\begin{equation}
\label{eq:passcal}
\frac{\partial T}{\partial t} 
+ \frac{\partial u_j T}{\partial x_j}
 =
 \alpha
 \frac {\partial^2 T}{\partial {x_j}^2}  
 + Q_\textrm{IBM} \mbox ,
\end{equation}
where $\alpha$ denotes the thermal diffusivity.
Periodic boundary conditions are applied for the thermal field in $x$-direction, while  constant temperature is prescribed on both the lower and upper walls of the flow domain.
The non-dimensionalized temperature is defined as $\theta=(T-T_l)/\Delta T_w$ with $\Delta T_w = T_u - T_l$, such that $\theta_l=0$ and 
$\theta_u=1$.
The Prandtl number is chosen to be $\mathrm{Pr}=\nu/\alpha=1$. 
$Q_\textrm{IBM}$ is proportional to the heat transfer rate between the flow and the structured wall and can be considered as a counterpart to the volume force $F_{\textrm{IBM},i}$ in the momentum equation. This term is adjusted to fulfill the temperature boundary condition on the structured wall.
Due to the use of periodic boundary condition for temperature, the absolute value of the heat transfer rate on the two walls should be identical once the solution reaches the thermal equilibrium. 
For the same reason, the mean heat flux in the wall-normal direction is constant in the channel.
The present thermal boundary condition is chosen following other studies of heat transfer above structured walls \cite{Leonardi2015, Miyake2001, Nagano2004}. 

The solver implementation is based on a spectral solver for incompressible boundary layer flows~\cite{chevalier_2007}. 
The Navier-Stokes equations are numerically integrated using the velocity-vorticity formulation by a spectral method with Fourier decomposition in the horizontal directions and Chebyshev discretization in the wall-normal direction. For temporal advancement, the convection and viscous terms are discretized using the third-order Runge-Kutta and Crank-Nicolson methods, respectively.
The flow domain is discretized with $N_x \times N_y = 256 \times 129$ grid nodes, while the immersed boundary method is applied on the dealiased grid (3/2 rule) with $384 \times 129$ grid nodes.

\subsection{Performance indices}
\label{sec:performance_indices}
Contrary to the laminar flow in a flat channel no universal analytical solution can be derived for a channel with arbitrary structuring at both channel walls.
Utilizing the melt-down heights of the imposed structure for both walls ($h_u$, $h_l$) and splitting the flow into two halves based on the position of the maximal spatially averaged velocity denoted with $y_c$ (Fig.~\ref{fig:lamprof3}), the balance between pressure drop $P_x$ and the average effective wall shear stress $\tau_{eff}$ is given by
\begin{equation}
    \tau_{eff} = \frac{(\delta_l+\delta_u)}{2} P_x,
\end{equation}
where $\delta_u$ and $\delta_l$ define the upper and lower effective channel half heights with respect to $y_c$.
Based on the wall shear stress the mean drag coefficient is given as
\begin{equation}
 C_f = \frac{2 \tau_{eff}}{\rho {U^{eff}_b}^2},
\end{equation}
where the effective bulk mean velocity
\begin{equation}
U_b^{eff} =
 \frac{1}{(\delta_u+\delta_l)} \int_0^{2\delta} \left< u \right> \mathrm{d}y = \frac{2 \delta}{(\delta_u+\delta_l)} U_b.
\end{equation}
The brackets $\left< \right>$ denote a quantity averaged in $x$-direction so a split-up into the mean part $\left< \phi \right> (y)$ and spatial fluctuation part $\phi^\prime (x,y)$ can be performed for any quantity $\phi(x,y)$:
\begin{equation}
    \phi(x,y) = \left< \phi \right> (y) + \phi^\prime (x,y).
\end{equation}

Due to the asymmetry in the temperature boundary condition, the heat transfer properties have to be separately evaluated for each wall. 
Hence, the hydraulic diameter is defined for the upper and lower walls as
\begin{equation}
    D_{h,u/l} = 4 \delta_{u/l} .
\end{equation}
The Nusselt number for both walls can be estimated with
\begin{equation}
    Nu_{u/l} = \frac{4 \delta_{u/l} q_{tot}}{\lambda \Delta \theta_{b,u}},
\end{equation}
where $q_{tot}$ denotes the total heat flux and the bulk mean temperature differences are defined as
\begin{equation}
   \Delta \theta_{b,l} = \frac{1}{\delta_l U_{b,l}^{eff}} \int_0^{y_c} \left< u \right> \left< \theta \right> \mathrm{d}y,
\end{equation}
and
\begin{equation}
   \Delta \theta_{b,u} = \frac{1}{\delta_u U_{b,u}^{eff}} \int_{y_c}^{2\delta} \left< u \right> (1-\left< \theta \right>) \mathrm{d}y.
\end{equation}
The average of $\mathrm{Nu}_{l}$ and  $\mathrm{Nu}_{u}$ is computed to determine the resultant Nusselt number of a particular case.
The effective bulk mean velocity for each channel half is given by
\begin{equation}
U_{b,l}^{eff} = \frac{1}{\delta_l} \int_0^{y_c} \left< u \right> \mathrm{d}y \quad \textrm{or} \quad U_{b,u}^{eff} = \frac{1}{\delta_u} \int_{y_c}^{2\delta} \left< u \right> \mathrm{d}y.
\end{equation}
The total heat flux $q_{tot}$, which is a constant as mentioned previously, can be estimated as the sum of
\begin{equation}
q_{tot} = \lambda  \frac{\mathrm{d} \left< {\theta} \right>}{\mathrm{d}y}  -\rho c_p  \left<{v^\prime \theta^\prime} \right>  + \rho c_p Q_\textrm{IBM}^y,
\end{equation}
where the three terms are respectively named the laminar, total fluctuation, and IBM contributions~\cite{stroh2020secondary}. Here 
 $c_p$ denotes the specific heat capacity and $Q_\textrm{IBM}^y = - \int_y^\delta Q_\textrm{IBM} \mathrm{d}y$.
Finally, the Stanton number is defined based on $Re_{D_h} = 2 (\delta_l+\delta_u) U_b^{eff} / \nu$ and Prandtl number $Pr$:
\begin{equation}
    St = \frac{Nu}{Re_{D_h} Pr}.
\end{equation}
Reynolds analogy factor $RA$ relates Stanton number to the drag coefficient 
\begin{equation}
    RA = \frac{2 St}{C_f},
\end{equation}
and is used to evaluate the similarity between drag coefficient and heat transfer \cite{bons2005critical}. An increase in $RA$ highlights a stronger enhancement in heat transfer compared to that in the drag coefficient and hence is desirable in the design of an energy-efficient fluidic system.
It has to be noted that for the chosen boundary conditions $RA=0.533$ with $St=0.016$ and $C_f=0.06$ in the flat channel configuration. 


\subsection{Dataset and machine learning model}
\label{sec:dataset_and_ML}

To generate a diverse dataset of wall structuring, we used a random walk algorithm combined with spline interpolation and discretization on the simulation grid. Each wall structure consists of $n$ supporting points between a start and an end point at $x=0$ and $x=384$ with the same $y$-position for periodic boundary conditions. The $x$-coordinates of the supporting points are sampled from:
\begin{equation}
    x \sim \mathcal{N}(\mu=i \cdot \frac{384}{n}, \sigma=\sigma_x \cdot \frac{384}{n})
\end{equation}
$i$ is the $i^{th}$ supporting point in the interval $[1, n]$. $\sigma_x$ is varied according to table \ref{tab:generator_parameters}.

The $y$-coordinate at each respective $x$ is sampled from:
\begin{equation}
    y \sim \mathcal{N}(\mu = y', \sigma = \sigma_y \cdot \Delta)
\end{equation}
With $y'$ being the previous y position, $\sigma_y$ varied from table \ref{tab:generator_parameters} and $\Delta$ the available build space in $y$-direction.
A minimum of 50\% of the channel height is kept empty for the flow. To allow for larger meanders, the generation algorithm of the first wall surface can use the full 50\% of build space, so here $\Delta$ remains at constant $64.5$ at all $ x$ positions. For the second channel surface, $\Delta$ is adjusted according to the first channel surface.
The first y-coordinate at $x=0$ to initialize $y'$ is drawn from a uniform distribution in the interval $[0, \Delta]$.

The obtained supporting points are then interpolated with cubic Bézier curves. The distance of the control points from the supporting points is determined by the parameter $r$. For small $r$'s also the radius of the curves can get very small, resulting in sharp features. The parameter $a$ controls the smoothness of the curve. For $a=0$ the angle through one supporting point is determined by the mean of the directions to both neighbouring points. At higher $a$ the direction to one neighbouring point is weighted higher and hence the curve features stronger edges\footnote{
\url{https://github.com/aimat-lab/ChemEngML}, 
\url{https://stackoverflow.com/a/50751932}}.

\begin{table}[tb]
    \centering
    \caption{Parameters for the structure generation algorithm.}
    \label{tab:generator_parameters}
    \begin{tabular}{l l}
        \hline
        parameter & variations \\
        \hline
        $n$ & $[2, 3, 4, 5, 6, 7, 8, 9, 10]$ \\
        $\sigma_x$ & $[0.1, 0.2, 0.3, 0.4, 0.5, 0.6]$ \\
        $\sigma_y$ & $[0.05, 0.1, 0.15, 0.2]$ \\
        $r$ & $[0.0, 0.05, 0.1, 0.15, 0.2]$ \\
        $a$ & $[0.0, 0.05, 0.1, 0.15, 0.2]$ \\
        \lasthline
    \end{tabular}
\end{table}

Those obtained curves were then projected on the dealiased grid. While the $x$-coordinates where linearly spaced, the $y$-coordinates of the grid nodes were obtained using:
\begin{equation}
    y = \left( -cos \left( i \cdot \frac{\pi}{128} \right) + 1  \right) \cdot \frac{1}{2 \cdot 129}
\end{equation}
With the integer $i$ in the interval $[0, 128]$. Using $20$ random initializations per parameter variation $108,000$ random wall structures were generated, hereafter called the {\it repository set}.

We calculate the drag coefficient $C_f$ and Stanton number $St$ using the simulation method described above for a subset of $10,800$ randomly sampled channel geometries.
From that, $9,185$ passed a set of filters regarding temperature convergence and geometric validity.
$5\,\%$ ($459$) of this data was put aside as the test set.

We used varying fractions of the remaining $8,726$ channel geometries (hereafter called the {\it labeled set}) for hyperparameter optimization of CNN. Hyperparameters were determined without flat channel geometries. The training and test sets include flat channels to boost geometry patterns. We used a total of $8,776$ channel geometries as the training set and $464$ channel geometries as the test set. 
The inputs for the CNN are the binary images, each with $384 \times 129$ pixels, representing a cross-section of the channel geometries, {\it i.e.} exactly the same input which is also used in the numerical simulations.
The input is passed through a varying {\it number of convolution steps}, each consisting of a convolution with padding, a varying {\it kernel size} and a varying {\it number of filters}, followed by a relu-activation and a $2$x$2$x max.-pooling with stride $2$.
The output of the convolutions is then flattened and passed through one hidden dense layer with relu-activation and a varying {\it number of neurons}.
This hidden layer is additionally regularized by a varying {\it dropout}.
$C_f$ and $St$ are then predicted with an output layer with two neurons and linear activation.
We trained the CNN model with augmented channel geometries to improve the model’s interpretability for different shapes of geometries. The final hyperparameter optimum was used to train the model. Data augmentation was implemented on each batch and epoch during training. To account for periodical boundary condition and invariance of vertical orientation the channel geometry was randomly flipped about the $y$-axis, and also randomly shifted in the $x$-dimension within the range of $x =$ 0 to 384. 

The model was implemented using TensorFlow \cite{tensorflow2015-whitepaper} and Keras \cite{chollet2015keras}, and training was done using the Adam optimizer \cite{kingma2017adam}.
The model is trained for 100 epochs with a batch size of $256$ using the mean squared error (MSE) as loss while logarithmically reducing the learning rate from $1e^{-3}$ at the $10^{th}$ epoch to a varying {\it final learning rate} at the last epoch.

To find out whether the prediction of $C_f$ and $St$ can be improved jointly this search space (table \ref{tab:hypopt_parameters}) is optimized using multi-metric Bayesian optimization. For this SigOpt is employed \cite{sigopt-paper, sigopt-docs} with $140$ experiments on $8$ asynchronous parallel channels. Each channel has access to a single Tesla A100 GPU.
To ensure stable predictions, for each evaluation the mean MSEs for the predicted $C_f$ and $St$ of a 5-fold cross-validation are used as metrics.

\begin{table}[tb]%
    \centering
    \caption{Parameters and final optimum for the hyperparameter optimization.}
    \begin{tabular}{l l l l l}
        \hline
        name & type & min & max & optimum\\
        \hline
        kernel size & int & $2$ & $20$ & \textbf{$7$} \\
        number of convolution steps & int & $2$ & $5$ & \textbf{$5$} \\
        number of  filters & int & $1$ & $128$ & \textbf{$128$} \\
        number of neurons & int & $50$ & $6000$ & \textbf{$6000$} \\
        final learning rate & double & $1.0e^{-11}$ & $1.0e^{-4}$ & \textbf{$6.105e^{-6}$} \\
        dropout & double & $0.05$ & $0.5$ & \textbf{$0.3$} \\
        \lasthline
    \end{tabular}
    \label{tab:hypopt_parameters}
\end{table}

\section{Results}
\label{sec:results}

\begin{figure*}[!htb]
  \centering
   \includegraphics[width=0.8\textwidth]{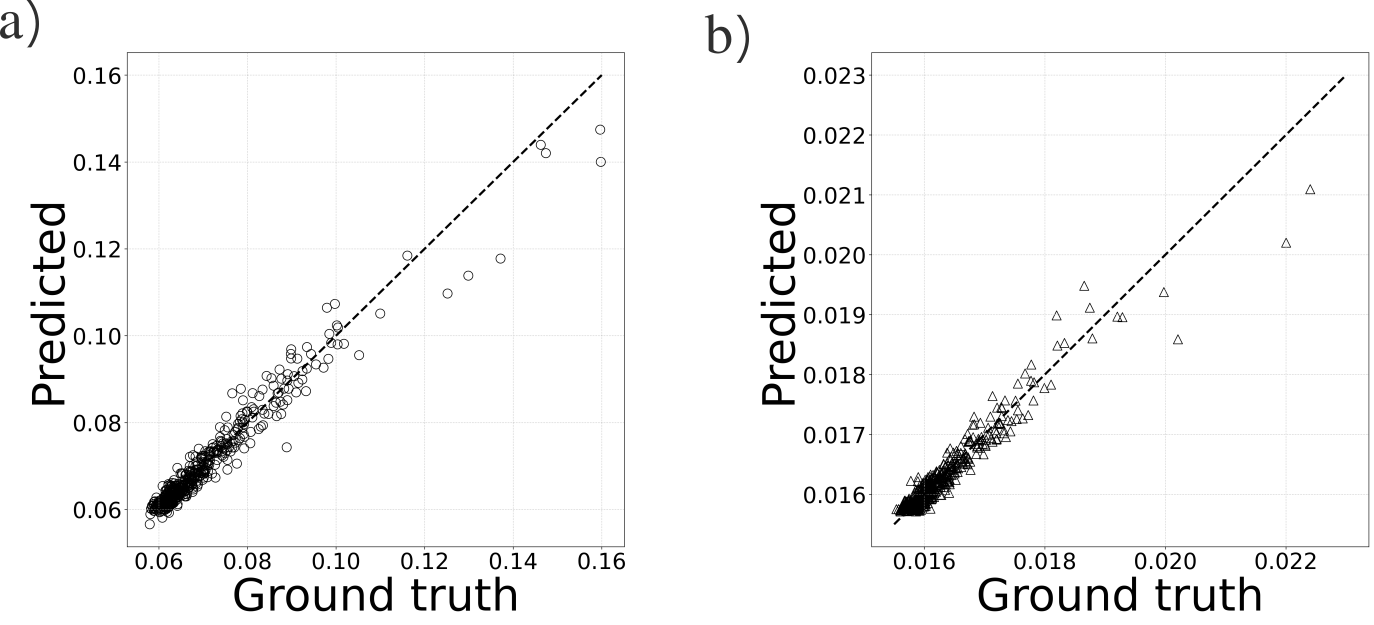}
   \caption{Predictions of the CNN model of a) drag coefficient $C_f$ and b) Stanton number $St$ compared to the ground truth on the validation set.}
   \label{fig:predictions_scatter}
 \end{figure*}

We trained the CNN model described above and obtained MAE (mean absolute errors) (and $r^2$-scores) of $MAE=1.90\cdot10^{-3}$ ($r^2=0.951$) and $MAE=1.29\cdot10^{-4}$ ($r^2=0.925$) for predictions of $C_f$ and ${St}$. 
A comparison of CNN predictions and simulated ground truth on the test set is shown in Figure~\ref{fig:predictions_scatter}. Most values of $C_f$ and ${St}$ are well correlated. However, deviation for large values can be seen($C_f$ = $0.12\sim$, ${St}$ = $0.020\sim$).

\begin{figure*}[!htb]
  \centering
   \includegraphics[width=0.8\textwidth]{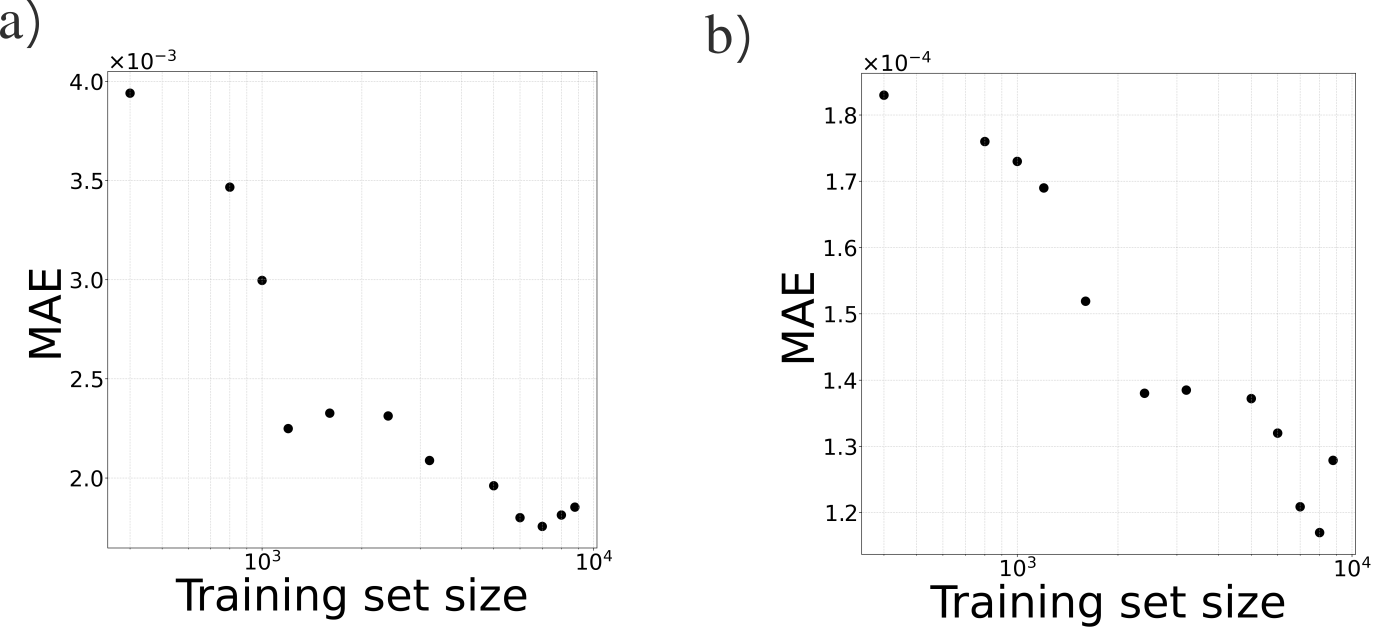}
   \caption{Learning curve, {\it i.e.} mean absolute error as a function of the training data size of 
 CNN model for a) drag coefficient $C_f$ and b) Stanton number $St$.}
   \label{fig:mae}
 \end{figure*}

In order to evaluate how well the CNN performs on smaller datasets, we generated learning curves (see Figure~\ref{fig:mae}), where we observe the MAE in $C_f$ and $St$ as a function of the training set size. The hyperparameters were kept constant, and the amount of training data was varied from $5~\%$ to $90~\%$ of a total of training and test set, 9,240 channel geometries. We trained the CNN model with each data set and evaluated the model performance by plotting mean absolute errors. We observe an exponential decrease in the mean absolute error with the increase in training set size. It can be observed from Figure~\ref{fig:mae} that larger datasets increase the model accuracy. In $C_f$ case, the MAE score for all data is approximately twice smaller than that of $5~\%$, meaning a large number of geometry channels is necessary to predict $C_f$ and $St$ with high accuracy. In $St$ case, a linear decrease as the size of the dataset is larger can be seen from the plot. As the plots continue to decrease exponentially, further improvement in the accuracy of the model can be expected by increasing the number of data.

\begin{figure*}[!htb]
  \begin{subfigure}{0.48\textwidth}
  \centering
  \includegraphics[width=0.85\textwidth]{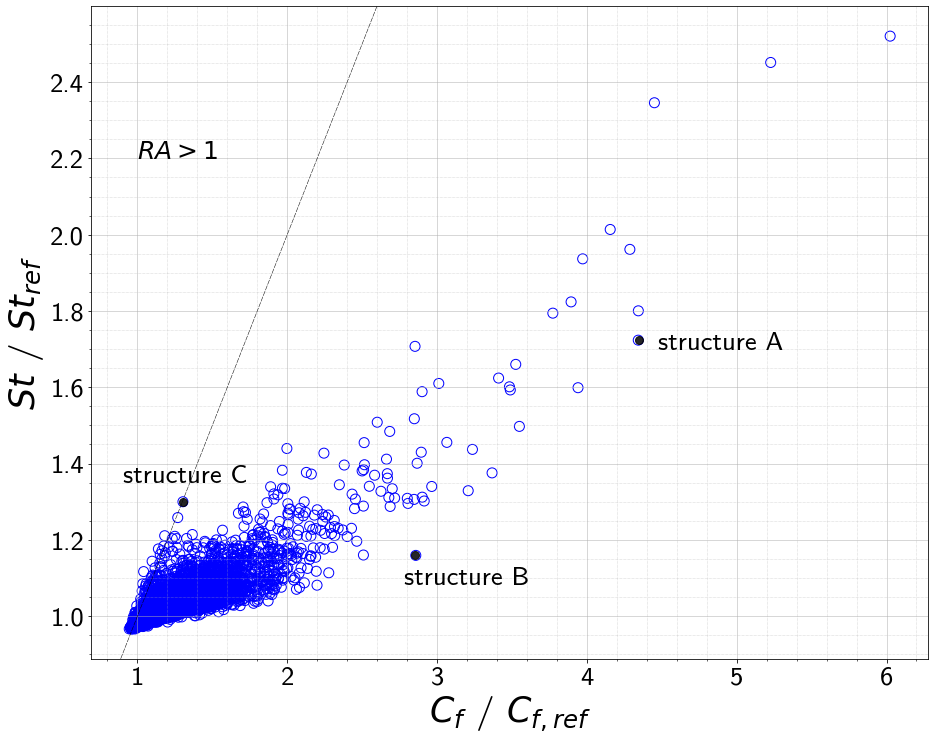}
   \caption{$C_f$ vs $St$ for the training set}
   \label{fig:scatterPlot_training}
  \end{subfigure}
   \begin{subfigure}{0.48\textwidth}
  \centering
   \includegraphics[width=0.85\textwidth]{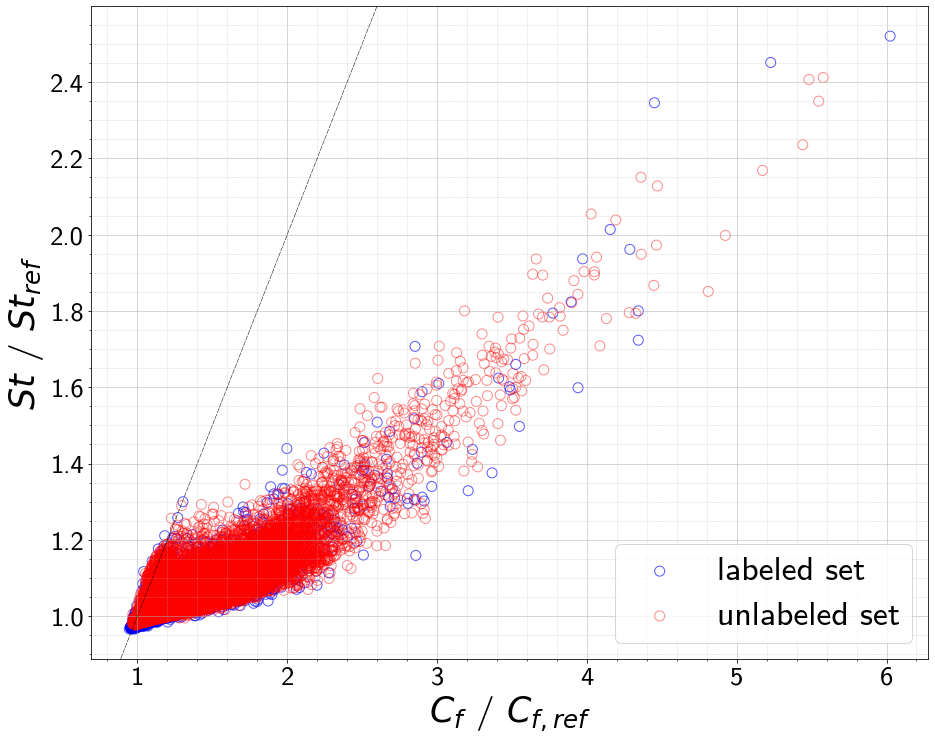}
   \caption{$C_f$ vs $St$ for the repository set}
   \label{fig:scatterPlot_sample_all}
  \end{subfigure}
   \begin{subfigure}{0.485\textwidth}
  \centering
   \includegraphics[width=0.9\textwidth]{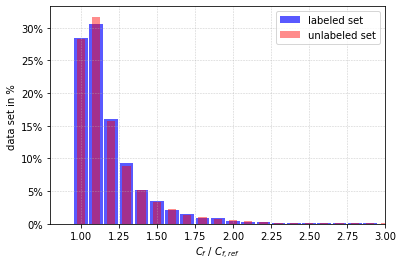}
   \caption{Histogram representation of $C_f$}
   \label{fig:combinedHistogram_Cf}
  \end{subfigure}
   \begin{subfigure}{0.485\textwidth}
  \centering
   \includegraphics[width=0.9\textwidth]{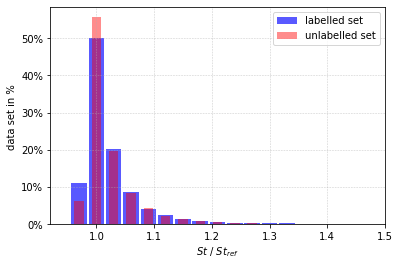}
   \caption{Histogram representation of $St$}
   \label{fig:combinedHistogram_St}
  \end{subfigure}
  \caption{Scatter plot of $C_f$ vs $St$ for the a) training (labeled) and b) repository set. Histogram representation of the training and repository set for c) $C_f$ and d) $St$.}
 \label{fig:Cf_St}
 \end{figure*}

 The scatter plot of $8,726$ labeled data points from the repository set used for training the CNN is shown in Figure~\ref{fig:scatterPlot_training}. Deviations from a flat channel ($C_{f,ref}=0.06$ and $St_{ref}=0.016$) necessarily lead to an increase in $C_f$ that outweighs the simultaneous increase in $St$. We would like to point out that the training set contains geometries that are well representative of the flow configurations usually encountered in fluid dynamics. Three of these structures that are highlighted in Figure~\ref{fig:scatterPlot_training} are discussed in detail in Section \ref{sec:Discussions}. We then exploited the speed up of the surrogate machine learning model ($<100~\mathrm{ms}$ per channel) compared to the numerical simulation ($\approx20-30~\mathrm{min}$ per channel). This allowed us to explore the flow and heat transfer characteristics ($C_f$ and $St$) of a much larger set of unlabeled channel geometries from the repository set. A scatter plot of $C_f$ vs $St$ for the repository set, split up into labeled and unlabeled data points, is shown in Figure~\ref{fig:scatterPlot_sample_all}. In general, it is very difficult to find geometries with $RA$ value exceeding that of a flat channel as can be seen in Figures~\ref{fig:scatterPlot_training} and \ref{fig:scatterPlot_sample_all}. The histogram representation of the repository set shown in Figures~\ref{fig:combinedHistogram_Cf} and \ref{fig:combinedHistogram_St} shows that the training data adequately represents the repository set.
 

\section{Discussions}
\label{sec:Discussions}

In this section, we analyze in detail the pressure loss and heat transfer characteristics of the three {\it labeled} structures highlighted in Figure~\ref{fig:Cf_St}a (namely structures A, B, and C). The mean flow and temperature fields of these structures are shown in Figure~\ref{fig:selectedStructures}, and their $C_f$, $St$, and $RA$ values are listed in Table~\ref{tab:CfSt_values}. Interestingly, these three structures resemble three canonical flow configurations encountered in fluid dynamics: vortex generator, converging-diverging nozzle, and backward-facing step. For the three considered structures, the percentage increase in surface area is $25.1\%$, $5.4\%$, and $6.3\%$ respectively.

Among the three structures, heat transfer achieved is maximum for structure A with the vortex-generating wing-like protrusion. Vortex generators of varying shapes, similar to the wing-like protrusion of structure-A, are commonly used in heat exchanger devices to introduce unsteady swirling motions that can increase heat transfer \cite{Fiebig1995}. Though there are two vortices inside the flow field as can be seen in Figure~\ref{fig:selectedStructures}a, these vortices are part of the recirculation regions. Such regions in fact isolate the wall from the bulk of the fluid and are detrimental to effective heat transfer. Nonetheless, the developing thermal boundary layers generated on both the top and bottom walls (refer to the temperature field in Figure~\ref{fig:selectedStructures}b) result in an 80\% increase in heat transfer. At the same time, the shape of the structure-A together with its increased surface area introduces significant pressure loss leading to a 60\% reduction in $RA$ when compared to the flat channel. 

For the structure B, flow inside  gets accelerated inside the converging section followed by fluid deceleration inside the diverging section. Unlike the case of structure A, a developing thermal boundary layer is present only along the bottom wall. The presence of the recirculation region, covering the entire bottom wall except for the converging-diverging section, further limits the heat transfer from the bottom wall. As a result, the increase in heat transfer is only 16\%. The considerable pressure loss due to the absence of a well-streamlined converging-diverging section leads to a 60\% reduction in $RA$ with respect to the flat channel. 

Structure C is unique owing to the fact that it contains the backward-facing step together with wall meandering. The structure results in a $RA$ value approximately equal to unity but with a 30\% increase in heat transfer with respect to the flat channel. This means that the structure results in a proportional increase in pressure loss as well. With the exception of the immediate area downstream of the backward step, the streamlines in the wall-normal direction exhibit behavior akin to those found within a rectilinear channel, maintaining a uniform and evenly spaced distribution. The absence of a developing thermal boundary layer indicates that wall meandering with a 6\% increase in the surface area should be the primary reason for the increased heat transfer. This suggests the possibility to achieve a value greater than unity for the $RA$ value with carefully selected parameters for wall meandering \cite{floryan2015flow}.


\begin{figure*}[t]
    \centering
    \input{tikz/pics_2d}
    \caption{Mean velocity and temperature fields for the structures A, B, and C. Streamlines are added on top of the velocity fields for a better understanding of the flow field, especially the recirculation regions. The white dashed lines in the temperature fields indicate $\theta = 0.10$ and $0.90$ and are representative of the thermal boundary layer.}
    \label{fig:selectedStructures}
\end{figure*}

\begin{table}[tb]
    \centering
    \caption{$C_f$, $St$, and $RA$ values for the structures A, B, and C. Also shown are the values normalized with the corresponding values for the flat channel, $i.e.$  ${C_f}_{ref} = 0.06$ and $St_{ref} = 0.016$.}
    \begin{tabular}{c||c|c||c|c||c|c}
    \bottomrule
         stru.              & $C_f$      & $\frac{C_f}{{C_f}_{ref}}$  & $St$    &$\frac{St}{St_{ref}}$  & $RA$ & $\frac{RA}{RA_{ref}}$ \\
         \hline
         A  &	0.2606	&	4.3429 &	0.0276	&	1.7875 &	0.2113 &	0.3965 \\
         B  &	0.1715	&	2.8580 &	0.01855	&	1.1593 &	0.2163 &	0.4058 \\
         C  &	0.0782	&	1.3037 &	0.0208	&	1.3006 &	0.5321 &	0.9976 \\
         \toprule
    \end{tabular}
\label{tab:CfSt_values}
\end{table}

\section{Conclusions and outlook}
 We presented a combination of accurate numerical simulations of fluid flow and heat transfer in arbitrary, non-flat channels and machine learning models predicting drag coefficient $C_f$ and Stanton number $St$. We found that $C_f$ and $St$ are well predicted from channel geometries by the CNN model with data augmentation. However, prediction is limited for the complex geometries. The higher numbers of $C_f$ and $St$ are difficult to predict the ground truth numbers. 
 We show that once trained the CNNs can predict the target properties at a fraction of the time ($<100~\mathrm{ms}$ per channel) required by numerical simulations ($\approx20-30~\mathrm{min}$ per channel). This can be exploited for exploration and optimization tasks \cite{IHTC2023}. 
 
 The general approach is not only applicable to simple flow setups as presented here but can be extended to more complex tasks, such as three-dimensional multiphase or even reactive unit operations in chemical engineering. The limitation will be the availability of data or the associated computational cost of the underlying simulations. Since the current CNN is trained for a specific set of flow conditions, in terms of Reynolds number and boundary conditions, it will be interesting to know how to modify the current model to cover a wide range of these conditions.

In order to further exploit ML models in general and CNNs in particular for the design of chemical engineering unit operations, we plan to implement active learning approaches and generative models to reliably explore the possible design space of channel structures and directly solve the inverse problem, {\it i.e.} the suggestion of channel architectures given desired target properties. Also, trying to predict not only the values of $C_f$ and $St$ but also the velocity and temperature fields inside these arbitrary geometries, for example using a physics-informed neural network, can be thought of as a natural extension of the present work.

\section*{Code and data availablity}
The code to train the CNNs can be found at https://github.com/aimat-lab/ChemEngML. The data that support the findings of this study are available upon reasonable request from the authors.

\section*{Acknowledgements}
This work was funded by the Deutsche Forschungsgemeinschaft (DFG, German Research Foundation) - FR 4072/3 - within the Priority Programme "SPP 2331: Machine Learning in Chemical Engineering". The authors acknowledge support from the state of Baden-Württemberg through bwHPC. P.F. has received funding from the European Union’s Horizon 2020 research and innovation programme under the Marie Sklodowska-Curie grant agreement No 795206.

\section*{References}
\bibliographystyle{plain} 
\bibliography{references}

\newpage


\end{document}